# Self-trapped excitons in solid Kr: Time resolved study

*A.N. Ogurtsov,[1,2] E.V. Savchenko,[1] S. Vielhauer[3] and G. Zimmerer[3]*

[1]Institute for Low Temperature Physics and Engineering of NASU, Lenin Avenue 47, 61103 Kharkov, Ukraine
[2]National Technical University "KhPI", Frunse Street 21, 61002 Kharkov, Ukraine
[3]Institut für Experimentalphysik der Universität Hamburg, Luruper Chausee 149, 22761 Hamburg, Germany

The relaxation of electronic excitations in rare gas solids is accompanied by branching of relaxation paths. In the course of relaxation an exciton can persist in a free state up to the moment of radiative transition to the ground state. In solid Kr the excitons have a predominant probability of being either self-trapped in perfect lattice or localized on lattice defects into excimer-like molecular centers. In recent studies there was suggested the existence of two types of molecular excited centers in solid Kr [1]. Radiative decay of these centers produces two molecular subbands $M_1$ and $M_2$. The excitation spectra of $M_1$ subband exhibit preferential excitation at photon energies below the bottom of the lowest $\Gamma(3/2)$, n=1 excitonic band.

On the other side, the spectra of photon yield at the edge of exciton absorption [2] showed the threshold energy $E_1$ of population of molecular trapped centers, which separate the range of photoexcitation where free and self-trapped excitons coexist from the range where photon absorption creates only free excitons. One of the possible process which may explain the sharp rise the excitation spectrum of $M_1$ subband at $E_1$ is the process of trapped center creation as a result of formation and nucleation states responsible for electronic excitation localization due to thermal lattice disorder in the ground state [3]. The measurements of the decays of $M_1$ and $M_2$ subbands of solid Xe revealed the difference in lifetime of the triplet states [4]. In the present paper we extend these studies on the case of solid Kr.

The experiments were performed at the SUPERLUMI experimental station at HASYLAB, DESY, Hamburg. Selective photon excitation was performed with $\Delta\lambda$=0.2 nm. The luminescence was spectrally dispersed by 0.5 m Pouey monochromator with $\Delta\lambda$=2 nm equipped with multisphere plate detector. If necessary, the time-window technique was used and signal was measured within a time window (length $\Delta t$) correlated with the excitation pulse (delayed by $\delta t$).

Figure 1 shows the emission (a) and excitation (c) spectra of free exciton (*FE*), $M_1$ and $M_2$ bands, example of decay of molecular subbands (b) and temperature dependence of triplet lifetimes of $M_1$ and $M_2$ bands (d). In addition to previously published data [5] we measured decay curves of both molecular subbands at excitation energies $E_1=E_{FE}$, $E_2$=10.59 eV, and $E_3$=11,07 eV denoted at Fig.1c by arrows. The decay curves were recorded at different temperatures at emission energies $h\nu_1$=8.18 eV ($M_1$) and $h\nu_2$=8.76 eV ($M_2$) denoted at Fig.1a by arrows.

Contrary to the case of solid Xe [4,5] there was no detectable difference in temperature dependence of "defect" $M_1$ subband under excitation by photons with $h\nu=E_{FE}$, and $h\nu>E_{FE}$. In the temperature range 10–50 K the temperature dependencies from Fig.1d may be well approximated by single exponent $\tau(T)=A\cdot\exp(B/T)$, where $A_1$=50 ns, $A_2$=59 ns, $B_1$=12.8 K, $B_2$=13.3 K for $M_1$ and $M_2$ subbands respectively. Thus, the time resolved spectroscopy unambiguously verified the existence of two types of molecular self-trapped excitons in solid Kr.

*Acknowledgements*. The support of the DFG grant 436 UKR 113/55/0 is gratefully acknowledged.

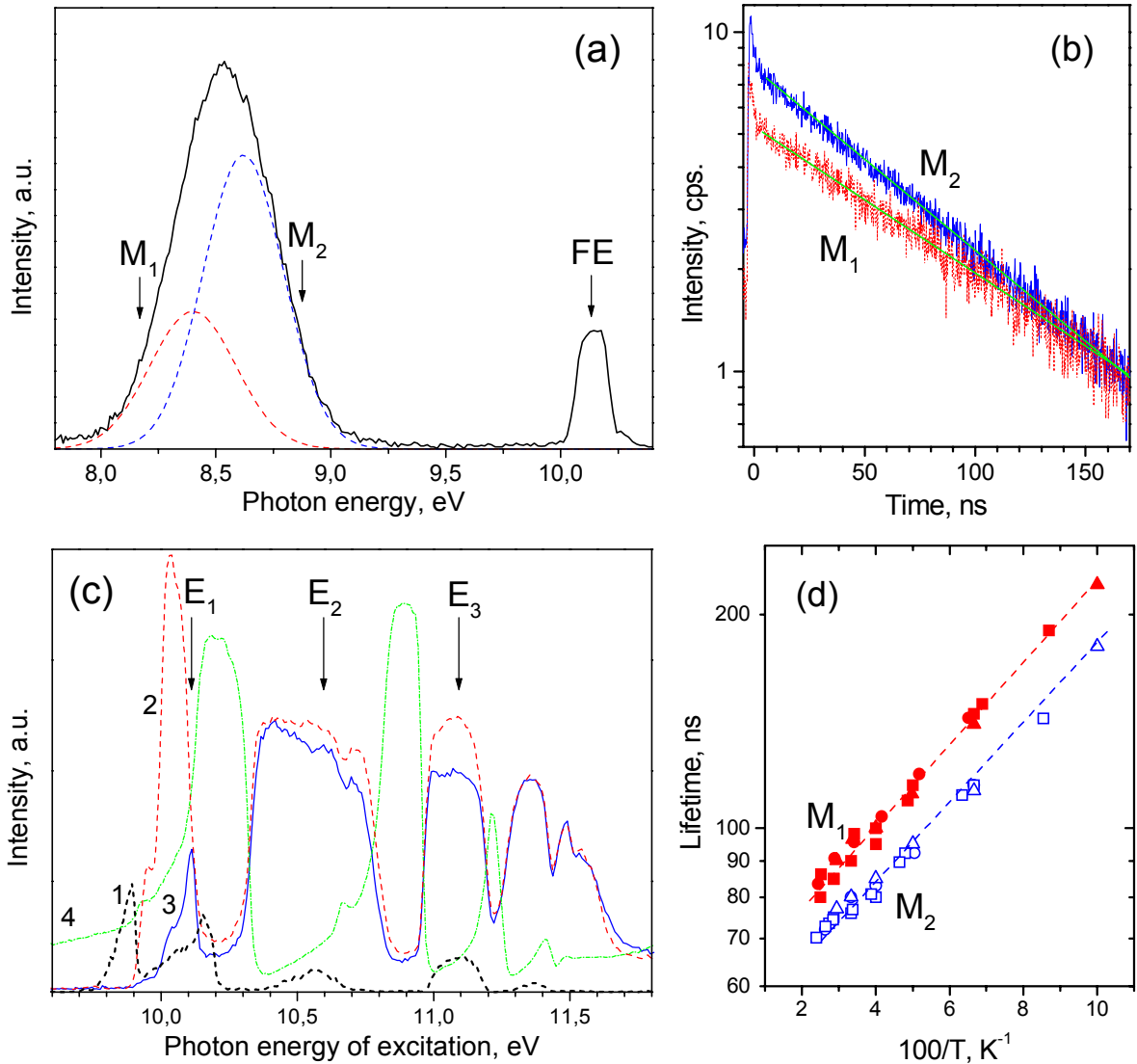

Figure 1: (a) – Luminescence spectrum of solid Kr under selective excitation by photons with $h\nu$=11.07 eV in time window with $\delta t$=1 ns and $\Delta t$=2 ns at T=30 K. (b) – Decay curves of molecular subbands measured at photon energies $h\nu_1$=8.18 eV ($M_1$) and $h\nu_2$=8.76 eV ($M_2$) (denoted at Fig.1a by arrows) under photoexcitation with energy $E_2$=10.59 eV. The curves were normalized at time $t$=160 ns. (c) – Excitation spectra of $M_1$ (curve 2), $M_2$ (curve 3) and *FE* (curve 1, enlarged 10 times) bands measured at photon energies denoted at Fig.1a by arrows, and reflection spectrum (curve 4) of solid Kr at $T$=25 K. (d) – Temperature dependence of the triplet luminescence of $M_1$ (solid red symbols) and $M_2$ (open blue symbols) molecular subbands of solid Kr under selective photoexcitation with energies $E_1$ (circles), $E_2$ (squares) and $E_3$ (triangles).

# References

[1] A.N. Ogurtsov, E.V. Savchenko, J. Low Temp. Phys. 122, 233 (2001).
[2] A.N. Ogurtsov, E.V. Savchenko, E. Gminder, S. Vielhauer, G. Zimmerer, Surf. Rev. Lett. 9, 45 (2002).
[3] J.L. Gavartin, A.L. Shluger, Phys. Rev. B 64, 245111 (2002).
[4] A.N. Ogurtsov, E.V. Savchenko, E. Sombrowski, S. Vielhauer, G. Zimmerer, Low Temp. Phys. 29, 858 (2003).
[5] A.N. Ogurtsov, E.V. Savchenko, S. Vielhauer, G. Zimmerer, HASYLAB Annual Report 2002 (DESY 2003) p.589.